\documentclass[11pt,twoside]{article}

\usepackage{asp2021}

\aspSuppressVolSlug
\resetcounters

\usepackage{xcolor}

\begin{document}

\title{The Astrosky Ecosystem: An independent online platform for science communication and social networking}

\author{
Emily L. Hunt,$^{1}$
Vincent S. Carpenter,$^2$
Kyle W. Cook,$^{3}$
Douglas G. Hilton,$^{4}$
Mehnaaz Asad,$^2$
Janine Jochum,$^{1}$
Kelly Lepo,$^{5}$
and Jamie Zvirzdin$^{6}$
}

\affil{$^1$Department of Astrophysics, University of Vienna, Türkenschanzstrasse 17, 1180 Wien, Austria; \email{emily.lauren.hunt@univie.ac.at}\\
$^2$Contributor, The Astrosky Ecosystem \\
$^3$Department of Physics \& Astronomy, University of Louisville, Natural Sciences Building 102, Louisville, KY 40208, USA \\
$^4$School of Earth and Space Exploration, Arizona State University, Tempe, AZ 85287, USA \\
$^5$Space Telescope Science Institute, 3700 San Martin Drive, Baltimore, MD 21218, USA \\
$^6$High-Energy Astrophysics Institute, Department of Physics \& Astronomy, University of Utah, Salt Lake City, UT 84112, USA}

\paperauthor{Emily L. Hunt}{emily.lauren.hunt@univie.ac.at}{0000-0002-5555-8058}{University of Vienna}{Department of Astrophysics}{Wien}{Wien}{1180}{Austria}
\paperauthor{Vincent S. Carpenter}{vincent.s.carpenter@gmail.com}{0000-0001-7919-2815}{The Astrosky Ecosystem}{}{Wien}{Wien}{1180}{Austria}
\paperauthor{Kyle W. Cook}{Kyle.cook@louisville.edu}{0000-0002-4012-779X}{University of Louisville}{Department of Physics \& Astronomy}{Louisville}{Kentucky}{40208}{USA}
\paperauthor{Douglas G Hilton}{astro.doug.hilton@gmail.com}{0009-0001-7438-7135}{Arizona State University}{School of Earth and Space Exploration}{Tempe}{Arizona}{85287}{USA}
\paperauthor{Mehnaaz Asad}{mehnaazasad93@gmail.com}{}{The Astrosky Ecosystem}{}{Wien}{Wien}{1180}{Austria}
\paperauthor{Janine Jochum}{janine.jochum@univie.ac.at}{}{University of Vienna}{Department of Astrophysics}{Wien}{Wien}{1180}{Austria}
\paperauthor{Kelly Lepo}{kellylepo@gmail.com}{0009-0008-0886-4168}{Space Telescope Science Institute}{}{Baltimore}{Maryland}{21218}{USA}
\paperauthor{Jamie Zvirzdin}{jamie.zvirzdin@utah.edu}{0009-0005-5418-0654}{University of Utah}{High-Energy Astrophysics Institute}{Salt Lake City}{Utah}{84112}{USA}



\begin{abstract}
While almost everything that astronomers study occurs in the vacuum of space, astronomy itself does not `happen in a vacuum'. Interactions between scientists, as well as outreach to members of the public, improve extensively from access to good communication tools. Social media has become a key tool for communication in astronomy, being widely used by individuals and organizations alike for networking, outreach, and more. However, traditional social media is reliant on benevolent corporations providing a free service without compromising on quality, and the recent takeover and decline of Twitter has shown how vulnerable these platforms can be. In this proceeding, we present The Astrosky Ecosystem, which is an initiative to develop open-source tools and integrations for social media, principally the Bluesky social network. We explain how our project enables the astronomy community to operate its own social media infrastructure, independent of for-profit corporations. We also discuss some of the project's technical aspects, including its use of the AT Protocol for social networking, before concluding with ideas for the future.
\end{abstract}




\section{Introduction}

Social media has revolutionized communication in astronomy. For outreach, social media offers an incredible opportunity to have fast, two-way communication with broad audiences. Varied uses of social media range in size from large organizations to individuals posting about their own work \citep{CominskyMcLin_2014,BlantonBershady_2017,RussoAlwast_2019,DalgleishProkoph_2022,Smethurst_2024}. For researchers, social media platforms offer a wide range of benefits, including as a tool for networking, discussing papers, or following conference news \citep{ShuaiPepe_2012,DarlingShiffman_2013,InclusiveAstronomy2_2020,WoodsWalker_2022}.

Owing to its real-time `microblogging' format, Twitter emerged as the main platform for discourse and outreach in astronomy. However, as `products' ran on a for-profit basis, traditional social media platforms are vulnerable to the whims of their owners. This fate unfortunately befell Twitter (now known as X) in 2022. Since then, X has plummeted in popularity across science because of a multitude of factors, including a sharp rise in hate speech and disinformation \citep{Mallapaty_2024,Kupferschmidt_2024}. The weakness of social media tools stands in stark contrast to the rest of the astronomical ecosystem: almost all other tools (such as the arXiv or Python packages) do not just disappear overnight. However, this is the current reality of the tools used for communication in astronomy, and the gradual decline of X has caused more than a decade of online astronomy community building to be lost.

In this work, we offer an alternative. Instead of relying on the whims of for-profit social media, we show that new social media technology can allow our community to host its own open-source social media infrastructure on a small budget. We discuss the Authenticated Transfer (AT) Protocol and how it enables this. Additionally, we present our experiences from the first 2.5 years of running The Astrosky Ecosystem online.

\section{The AT Protocol}

The core problem of running a social network is the need to ingest and distribute large amounts of data in real time. Traditionally, this has limited operating social networks to large corporations, who usually build bespoke, non-interoperable services that must be run on a for-profit basis to be financially sustainable. In this model, the corporation owning the platform has complete control over all content and data on it and can even sway algorithms when and how they please -- such as amplifying disinformation if it increases user engagement and hence advertising revenue \citep{aimeur_fake_2023}.

Modern, decentralized, and highly interoperable social media protocols like the AT Protocol offer an alternative to this model. \textcolor{blue}{\href{https://atproto.com/}{The AT Protocol}} \citep{KleppmannFrazee_2024} is an open-source protocol that splits traditional social media into multiple separate components. All data for a user across all AT Protocol applications is hosted on a Personal Data Server (PDS), with data types (such as posts on an app) following publishable schemas. Data from all PDSes is aggregated by a relay. Relays then distribute this data to other services, including moderation services, feed generation services, or applications. Finally, applications show a view of the network to users, and their interactions with it (such as `likes') are stored on their PDS and distributed around once again. All of these components can be hosted independently by any entity, and by virtue of extensive protocol design and optimization efforts, all components remain interoperable and cost-effective at scale \citep{KleppmannFrazee_2024}.

The most prolific example of an AT Protocol app is Bluesky. Bluesky is an open-source Twitter-like `microblogging' platform that has risen to prominence within science, becoming one of the de-facto replacements for Twitter \citep{Fieldhouse_2025}. By virtue of being an AT Protocol-based app, components of the platform (such as feed generators or PDS hosting) can already be run independently. In the event that the company operating Bluesky becomes an adversary, users would be able to migrate their data to new PDS hosts and pick up where they left off on a new clone of the original Bluesky app. Amongst a lively open-source AT Protocol developer community, multiple parties are already working on ensuring that this resilience exists within the network, such as the \textcolor{blue}{\href{https://blackskyweb.xyz/}{Blacksky}} and \textcolor{blue}{\href{https://www.eurosky.social/}{Eurosky}} initiatives.

\articlefigure[width=1.0\textwidth]{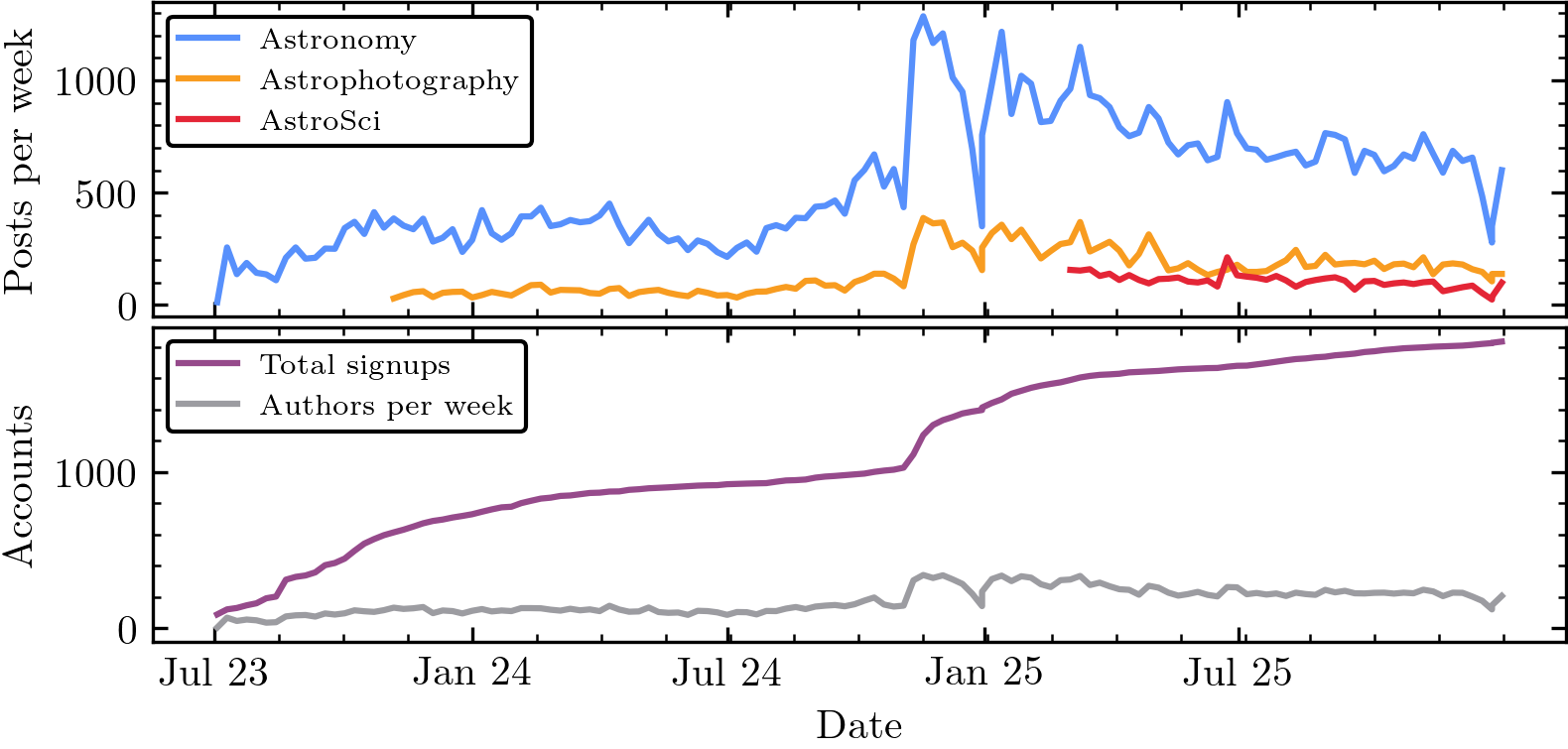}{posts}{Top: Number of posts per week to the Astronomy, Astrophotography, and AstroSci feeds made by registered posters. Bottom: Total number of registered posters on the feeds and the number of unique post authors per week.}

\section{The Astrosky Ecosystem}


Beginning in July 2023, our project has been curating an ecosystem of \textcolor{blue}{\href{https://github.com/the-astrosky-ecosystem}{open-source tools}}\footnote{\textcolor{blue}{\url{https://github.com/the-astrosky-ecosystem}}} within the AT Protocol space to empower astronomers and astronomy organizations online. We aim to not only replicate the mostly positive 2010s experience of Twitter for astronomy but even exceed it. Our main effort so far has been in producing and curating the Astronomy feeds, which are a network of feeds with posts about astronomy and astrophotography that can be \textcolor{blue}{\href{https://bsky.app/profile/did:plc:jcoy7v3a2t4rcfdh6i4kza25/feed/astro}{viewed on Bluesky}}. In total, we operate seventeen chronological feeds of posts, the three most popular of which currently see hundreds of posts per week from around 200--300 different users (Figure \ref{posts}).

The feeds have formed a central meeting point for the astronomy community and astronomy enthusiasts on Bluesky. Some of the largest accounts posting to the feeds include the European Space Agency, the Rubin Observatory, the European Southern Observatory, and the American Astronomical Society. In addition, the feeds include a diverse community of individual astronomers -- of which many registered posters are astrophotographers, who help the public engage with astronomy and whose eye-catching posts show that the night sky is observable by anyone \citep{SparksKruse_2019}.

Since March 2025, we have also been logging the number of requests to our webserver. Our server has seen an average of 740\,000 requests per week across the last ten months from an average of 92\,000 different accounts per week. Another metric for feed interaction is the number of times someone `scrolls' the feed (i.e., views more than the first $\sim${}20 posts): On average, the feeds are scrolled 91\,000 times per week by 19\,000 unique accounts. Posting to the feeds can massively elevate the reach of posts by astronomers and astrophotographers online, with even small accounts able to reach a wide audience. The number of viewers of the feeds far exceeds the total number of registered posters (1836 as of January 2026). 

These feeds, and all associated data to run them, are hosted independently by our project. All source code for the feeds is written in Python, chosen because of its familiarity to most astronomers, and the source code is available as \textcolor{blue}{\href{https://github.com/the-astrosky-ecosystem/astronomy-feeds}{an open-source library on GitHub}}. In addition, the feeds are moderated to be free of disinformation, including a sign-up step before users can post to the feeds. This ensures that only real astronomers or astrophotographers post to the feeds. We recently began \textcolor{blue}{\href{https://opencollective.com/the-astrosky-ecosystem}{a recurring crowdfunding campaign}}, which now financially covers our hosting costs (approximately €40 per month) in addition to some other project expenses.

In the near future, we are planning on expanding the range of the different tools we offer. Our next priority will be offering PDS (user data) hosting for astronomers online, cementing the independence and longevity of all astronomy content on AT Protocol applications. Better integrations with conferences, the arXiv, or telescopes are a possibility, particularly given the option for AT Protocol to quickly build new applications or data schemas while reusing existing account and data infrastructure. We are also investigating ways to further improve our feeds, such as by offering options to partially filter content by a user's primary interests or improving our registration process.

It is clear that the real-time, independent, and open-source functionality of the AT Protocol offers many advantages for social networking for astronomy. Many avenues to apply this new technology remain unexplored, and many more ideas could be explored in the future: for example, integrating the AT Protocol with existing publication infrastructure or even broadcasting detected astronomical transients over the AT Protocol. We welcome new contributions to our project, and we are excited to see where the AT Protocol may lead as we share and discuss astronomy online.

\acknowledgements We thank previous moderators and developers for their contributions to the project. We also thank the many developers in the lively and friendly open-source AT Protocol community for their support and encouragement. In particular, we thank Ilya Siamionau for developing the AT Protocol Python SDK, and for his help with bug fixes throughout the past 30 months. The authors thank Open Collective Europe for financially hosting the project.

\bibliography{O8-4}  


\end{document}